\documentclass[twocolumn,superscriptaddress,showpacs,preprintnumbers,amsmath
,amssymb,aps,prd]{revtex4}
\usepackage{graphicx}
\usepackage{slashbox}
\usepackage{bm}
\usepackage{color}
\usepackage{longtable}
\graphicspath{{Figures/}}

\def\cob#1{compact object binaries#1
  (COB#1)\gdef\cob{COB}}
\def\imbh#1{intermediate-mass black hole#1
  (IMBH#1)\gdef\imbh{IMBH}}
\def\bh#1{black hole#1
  (BH#1)\gdef\bh{BH}}
\def\bbh#1{binary black hole#1
  (BBH#1)\gdef\bbh{BBH}}
\def\gw#1{gravitational wave#1
  (GW#1)\gdef\gw{GW}}
\def\nr#1{numerical relativity#1
  (NR#1)\gdef\nr{NR}}
\def\snr#1{signal-to-noise-ratio#1
  (SNR#1)\gdef\snr{SNR}}
\def\pn#1{post-Newtonian#1
  (PN#1)\gdef\pn{PN}}
\def\GT#1{Georgia Tech#1
  (GT#1)\gdef\GT{GT}}

\def\newacronym#1#2#3{\gdef#1{#3 (#2)\gdef#1{#2}}}
\newacronym{\ligo}{LIGO}{Laser Interferometer Gravitational Wave Observatory}

\newcommand{\maya}{\textsc{Maya}}
\newcommand{\cactus}{\textsc{Cactus}}
\newcommand{\carpet}{\textsc{Carpet}}
\newcommand{\ET}{\textsc{EinsteinToolkit}}
\newcommand{\kranc}{\textsc{Kranc}}

\makeatletter
\let\protect\relax
{\catcode`\|=\active
  \xdef\InnerProduct{\protect\expandafter\noexpand\csname InnerProduct
\endcsname}
  \expandafter\gdef\csname InnerProduct \endcsname#1{%
    \begingroup
    \ifx\SavedDoubleVert\relax
    \let\SavedDoubleVert\|\let\|\IpDoubleVert
    \fi
    \mathcode`\|32768\let|\IPVert
    \left({#1}\right)
    \endgroup
  }
}
\def\IPVert{\@ifnextchar|{\|\@gobble}% turn || into \|
     {\egroup\,\mid@vertical\,\bgroup}}
\def\IPDoubleVert{\egroup\,\mid@dblvertical\,\bgroup}
\let\SavedDoubleVert\relax
\def\midvert{\egroup\mid\bgroup}
\def\SetVert{\@ifnextchar|{\|\@gobble}% turn || into \|
    {\egroup\;\mid@vertical\;\bgroup}}
\def\SetDoubleVert{\egroup\;\mid@dblvertical\;\bgroup}
\def\mid@vertical{\mskip1mu\vrule\mskip1mu}
\def\mid@dblvertical{\mskip1mu\vrule\mskip2.5mu\vrule\mskip1mu}
\makeatother

\usepackage{braket}

\begin{document}

\title{Georgia Tech Catalog of Gravitational Waveforms}
\author{Karan Jani}
\affiliation{Center for Relativistic Astrophysics and
School of Physics\\
Georgia Institute of Technology, Atlanta, GA 30332}
\author{James Healy}
\affiliation{Center for Computational Relativity and Gravitation \\ 
Rochester Institute of Technology, Rochester, NY 14623}
 \affiliation{Center for Relativistic Astrophysics and
School of Physics\\
Georgia Institute of Technology, Atlanta, GA 30332}
\author{James A. Clark}
\affiliation{Center for Relativistic Astrophysics and
School of Physics\\
Georgia Institute of Technology, Atlanta, GA 30332}
\author{Lionel London}
\affiliation{School of Physics and Astronomy, Cardiff University\\
 Queens Building, CF24 3AA, Cardiff, United Kingdom}
\affiliation{Center for Relativistic Astrophysics and
School of Physics\\
Georgia Institute of Technology, Atlanta, GA 30332}
\author{Pablo Laguna}
\affiliation{Center for Relativistic Astrophysics and
School of Physics\\
Georgia Institute of Technology, Atlanta, GA 30332}
\author{Deirdre Shoemaker}
\affiliation{Center for Relativistic Astrophysics and
School of Physics\\
Georgia Institute of Technology, Atlanta, GA 30332}

\begin{abstract} 
This paper introduces a catalog of gravitational waveforms from the bank of simulations by the numerical relativity effort at Georgia Tech. Currently, the catalog consists of 452 distinct waveforms from more than 600 binary black hole simulations: 128 of the waveforms are from binaries with black hole spins aligned with the orbital angular momentum, and 324 are from precessing binary black hole systems. The waveforms from binaries with non-spinning black holes have mass-ratios $q = m_1/m_2 \le 15$, and those with precessing, spinning black holes have $q \le 8$.  The waveforms expand a moderate number of orbits in the late inspiral, the burst during coalescence, and the ring-down of the final black hole. Examples of  waveforms in the catalog matched against the widely used approximate models are presented. In addition, predictions of the mass and spin of the final black hole by phenomenological fits are tested against the results from the simulation bank. The role of the catalog in interpreting the GW150914 event and future massive binary black-hole search in LIGO is discussed. The Georgia Tech catalog is publicly available at \href{url}{einstein.gatech.edu/catalog}.
\end{abstract}

% 04.25.D- Numerical relativity %
% 04.25.dg Numerical studies of black holes and black-hole binaries %
% 04.30.-w Gravitational waves 04.30.Db Wave generation and sources %
% 04.80.Nn Gravitational wave detectors and experiments

\maketitle

%%%%%%%%%%%%%%%%%%%%%%%%%%%%%%%%%%%%%%%%%%%%%
\section{Introduction}
\label{sec:intro}

Gravitational wave astronomy is finally here with the detection of transient GW150914~\cite{2016PhRvL.116f1102A}. The detection was both a  triumph and a surprise: a triumph because the \ligo{}~\cite{Harry:2010zz}  achieved unprecedented  sensitivity, and a surprise because of the particular characteristics of the source. The GW150914 transient was identified~\cite{2016arXiv160203840T} as the \gw{s} produced by the merger of a \bbh{}
% (see . \ref{f:bbh}) 
at a distance of $410^{+160}_{-180}$  Mpc. The masses of the \bh{s} were surprisingly large ($m_1 =  36^{+5}_{-4} \,M_\odot$ and $m_2 = 29^{+4}_{-4} \,M_\odot$, $q = m_1/m_2 \approx 1.22$) with net spins canceling each other ($\chi_{\text{eff}} \approx -0.06$). It is estimated that the coalescence left  behind a rotating \bh{} with a mass $M_f = 62^{+4}_{-4}\,M_\odot$ and spin $\chi_f  = 0.67^{+0.05}_{-0.07}$, thus suggesting that about $ 3\,M_\odot$ was emitted in \gw{s}. 

The role of  \nr{} simulations was evident in GW150914 event. The detection paper~\cite{2016PhRvL.116f1102A} showed the best fits of a \nr{} waveform to the data. The papers on parameter estimation~\cite{2016arXiv160203840T} and tests of general relativity~\cite{2016arXiv160203841T} made it clear that results from \bbh{} simulations were used to build the \texttt{SEOBNRv2}  and \texttt{IMRPhenomPv2} waveform models used in the analysis. And directly relevant to the present work, the paper on the analysis of the GW150914 event with minimal assumptions~\cite{2016arXiv160203843T} included results of matches using waveforms from the \GT{} catalog introduced in this paper.

\begin{figure*}[!htb]
  \includegraphics[scale=0.25,trim = 100 40 20 0]{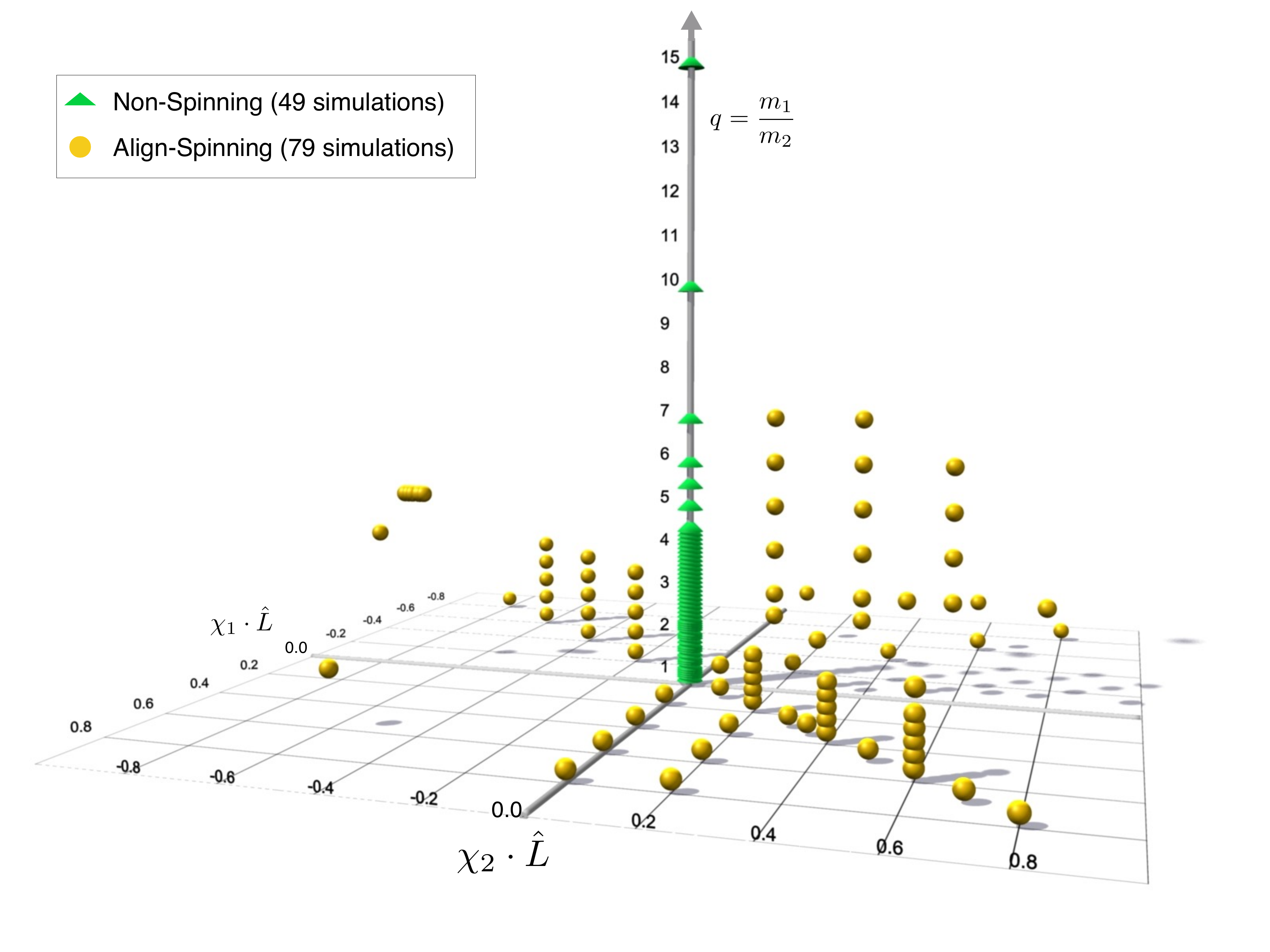}
  \includegraphics[scale=0.25,trim =0 40 90 0]{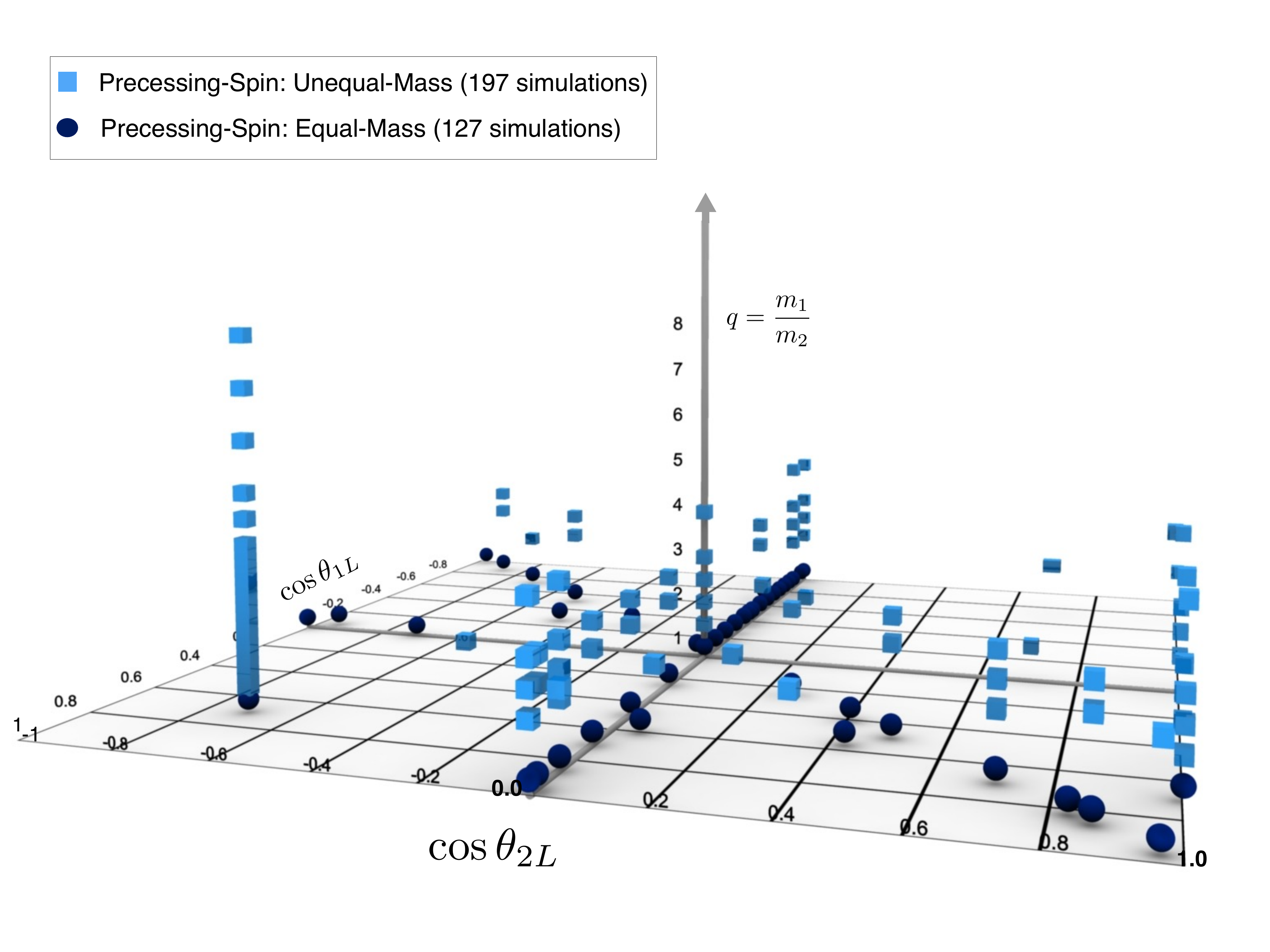}
  \caption{Coverage of binary black hole parameter space by the \GT{} catalog. The vertical axis in both plots denotes the mass ratio $q$. The plot on the left is for non-spinning and aligned-spin systems, and on the right for precessing binaries.}
\label{f:runs}
\end{figure*}

The goal of this paper is to formally introduce the \GT{} catalog of \gw{} waveforms. Currently, the catalog consists of 452 distinct waveforms from a bank of more than 600 \bbh{} simulations produced by the \nr{} effort at \GT{}. Among the 452 waveforms, 128 are from binary systems with \bh{s} non-precessing spins, i.e. no spins or spins such that they are parallel (aligned, or anti-aligned) with the orbital angular momentum $\vec L$); and, 324 waveforms are from generic spin configuration that lead to precessing \bbh{} systems (see Fig. \ref{f:runs}). The catalog probes mass-ratios of  $q\le15$ for binaries with non-spinning \bh{s} and $q\le8$ for binaries with precessing, spinning holes.  The waveforms cover  a moderate number of  \gw{} cycles in the late inspiral, the merger of the binary, and ends with the ring-down of the final \bh{}.

The waveforms are given in terms of an adjustable mass scale (the total mass $M = m_1+m_2$ of the \bbh{} system); and, therefore, they can be rescaled for both ground and space-based \gw{} detectors. In this paper, we focus the discussion on the relevance of the catalog to data analysis for ground detectors such as~\ligo{}.

Within the sensitivity window of~\ligo{} ($10-1000$ Hz) the waveforms in the catalog can be in general used in two ways. 
For binary systems with masses $M \ge 60\,M_\odot$, as in GW150914, the binary system is observed for less than half a dozen \gw{} cycles before merger. 
A substantial fraction of the waveforms in the \GT{} catalog expand this dynamical range. They can thus be applied directly in analysis massive \bbh{} mergers.
On the other hand, for binary systems with $M \le 60\,M_\odot$, more cycles are needed for detection and parameter estimation~\cite{abbott:2005:sfg:1,
abbott:2006:sfg:1,abbott:2008:sfg:1,abbott:2008:sos}. Our catalog also includes waveforms with enough cycles to help improve Effective One Body Approach (EOB)~\cite{PhysRevD.81.084041} and IMR (Inpiral-Merger-Ringdown)~\cite{Ajith:2009bn} waveform models.

The paper is organized as follows: Section~\ref{sec:code} provides a description of the \nr{} code used to produce the catalog, namely the \maya{} code. This Section also includes a discussion of the errors in phase and amplitude of the extracted \gw{s}. Section~\ref{sec:catalog} describes the parameter space and some of the key features of the \GT{} catalog.  Section~\ref{sec:EOB} compares a few of the waveforms in the catalog with  the\texttt{SEOBNRv2} and \texttt{IMRPhenomPv2} waveform models. Section~\ref{sec:fits} compares the parameters of remnant \bh{}, namely mass and spin, with the phenomenological fits~\citep{Barausse:2009uz, Barausse:2012qz, Healy:2014yta}. Conclusions are given in Sec.~\ref{sec:conclude}.

\begin{figure*}[htb!]
		\includegraphics[scale=0.43, trim = 450 30 300 50 ]{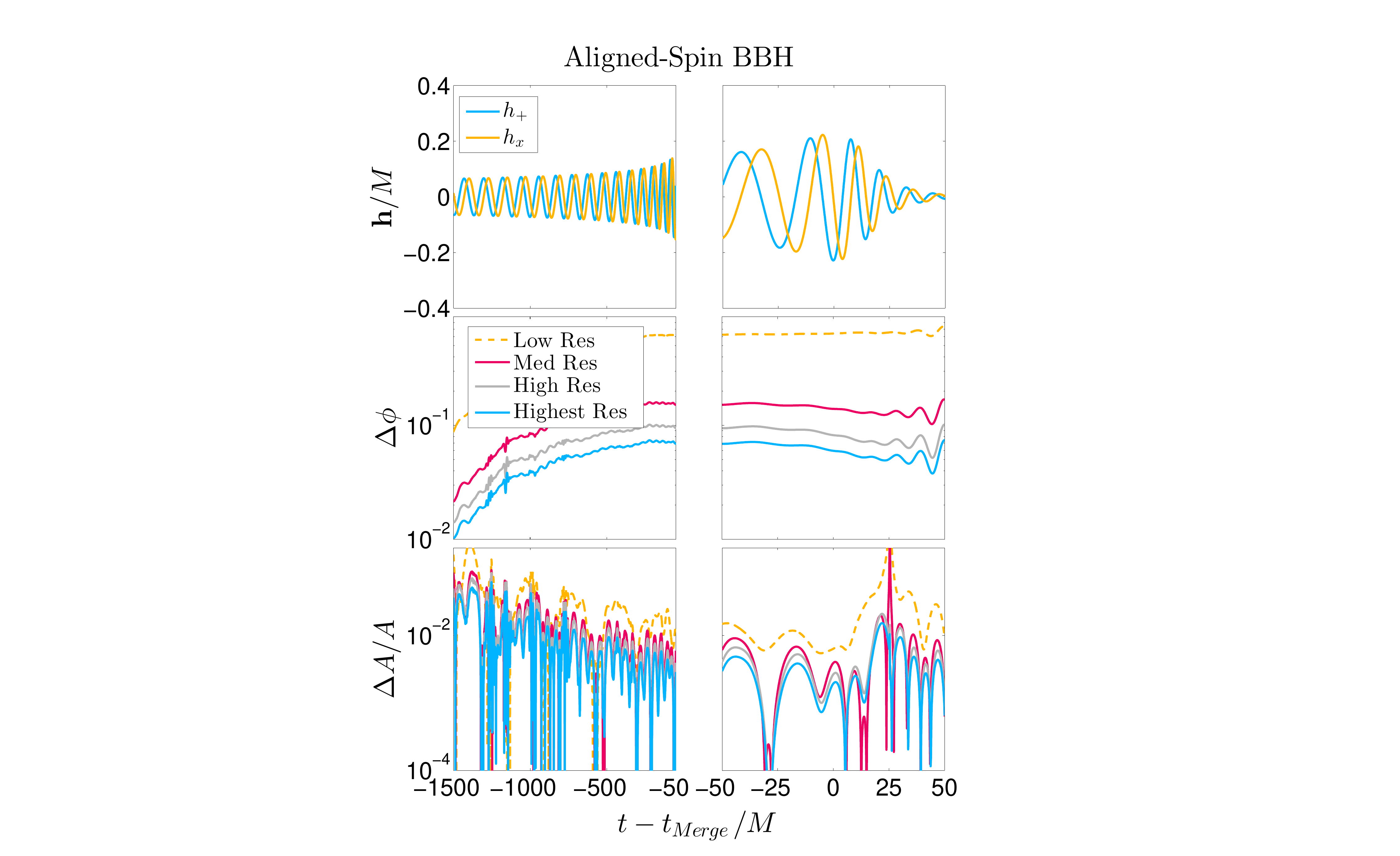}
				\includegraphics[scale=0.41, trim = 150 30 300 50]{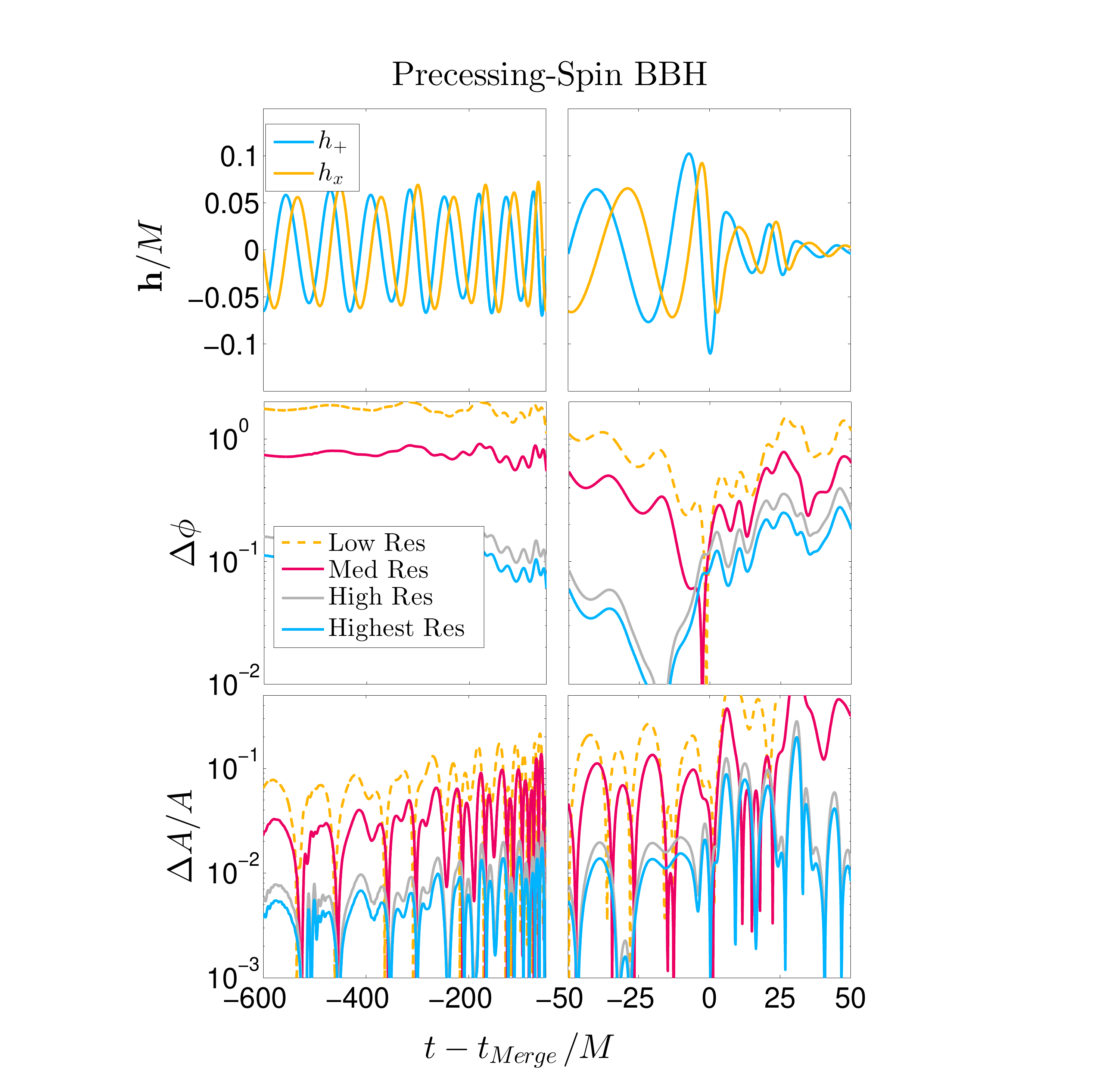}
\caption{Numerical errors in the amplitude and phase of the \gw{} strain, $h(t)$, for $\ell=2:6, m=-\ell:\ell$ radiated modes. Left panels show results for the  \texttt{GT0582} case and right panels for the \texttt{GT0560}. Top panels depicts the strain $h(t)$ at face-on location from detector. The middle and bottom panel shows the errors in phase and amplitude, respectively.
}
\label{f:err-analysis}
\end{figure*}

\begin{table*}[htb!]
  \begin{center}
  \begin{tabular}{p{1in} p{1.1in} p{1.1in} | p{0.8in} p{0.9in} p{0.8in} p{0.8in} }
   \multicolumn{1}{r}{} & \multicolumn{2}{l}{Mismatches from:} &  \multicolumn{2}{l}{Errors in GW-strain:} 	\vspace{0.05in}
\\
BBH Type & Finite Resolution & Finite Extraction & $\Delta A/ A |_{\text{Ins}}$ & $\Delta \phi |_{\text{Ins}}$& $\Delta A/ A |_{\text{Ref}}$  & $\Delta \phi |_{\text{Ref}}$	\\
  \hline\hline
 Aligned-Spin &  $3.4\times10^{-6}$ & $5.4\times10^{-5}$&  $5.8\times10^{-4}$ & $1.0\times10^{-2}$ & $3.7\times10^{-2}$ &$6.8\times10^{-2}$ \\
 Precessing-Spin  &  $4.0\times10^{-4}$ & $4.7\times10^{-4}$&  $3.6\times10^{-3}$ & $1.1\times10^{-1}$ & $1.2\times10^{-3}$ &$2.6\times10^{-2}$ 
\end{tabular}
\end{center}
  \caption{Typical numerical errors in GW strain for GT catalog. The numbers refer to the waveforms showcased in fig. \ref{f:err-analysis}.  }
  \label{tbl:err-analysis}
\end{table*}

\section{ Maya Code and Error Analysis}

\label{sec:code}
All the  \bbh{} simulations in the \GT{} catalog were obtained with our \maya{} code~\cite{Herrmann:2006ks, Vaishnav:2007nm, Healy:2009zm, Pekowsky:2013ska}. The code is based on the BSSN formulation of the Einstein equations~\cite{Baumgarte99}, and for \bbh{} simulation it uses the moving puncture gauge condition~\cite{Campanelli:2005dd,Baker:2005vv}.  \maya{}  is very similar to the Einstein code in the \ET{} \cite{et-web}.
That is, it operates under the  \cactus{} infrastructure~\cite{Allen99a}, with \carpet{} providing mesh
refinements~\cite{Schnetter-etal-03b} and thorns (modules) generated by
the package \kranc{}~\cite{Husa:2004ip}. 

The initial data for each simulation consist of the extrinsic curvature and spatial metric. The extrinsic curvature has the Bowen-York~\cite{Bowen80} form, and the spatial metric is conformally flat. The conformal factor is obtained by solving the Hamiltonian constrain using the \textsc{TwoPunctures} spectral solver~\cite{Ansorg:2004ds}.  The input parameters for each initial data set are:  \bh{} masses $m_1$, $m_2$, spins $\vec \chi_1$, $\vec\chi_2$, momenta $\vec P_1$, $\vec P_2$, and the binary separation $r$. A script that solves the \pn{} equations of motion for binaries in quasi-circular 
orbits~\cite{Husa:2007rh,Tichy:2010qa} is used to set  the spins $\vec \chi_1$, $\vec\chi_2$ and momenta $\vec P_1$, $\vec P_2$ at the binary separation $r$ in the initial data where the \nr{} evolution will start. The  mass and  spin of the final \bh{} are obtained from both its apparent horizon and quasi-normal ringing.

The \gw{} waveforms are extracted from the simulation data via the Weyl Scalar $\Psi_4$~\cite{gr-nr-methods-ExtractionAtInfinity-Reisswig2009}. 
The extraction is done in the source frame such that the initial orbital angular momentum of the binary is pointing in the positive $z$-direction.
We store $\Psi_4$ decomposed into spin-weighted spherical harmonics as
\begin{equation}
\label{eq:Psi4Decom}
R\,M\,\Psi_4(t;\Theta,\Phi) = \sum_{\ell,m}A_{\ell m}(t)e^{i\phi_{\ell m}(t)}\, {}_{-2} Y_{\ell m}(\Theta,\Phi) \,,
\end{equation}
with both $A_{\ell m}$ and $\phi_{\ell m}$ real functions of time, $M$ the total mass of the binary, and $R$ the extraction radius.   
Given $\Psi_4$, the \gw{} strain polarizations $h_+$ and $h_\times$ are obtained from integrating $\Psi_4 = \ddot h_+ - i\,\ddot h_\times \equiv \ddot h^\star$, with star denoting complex conjugation and over-dots time derivatives.  

To give a general sense of the accuracy of the waveforms, 
we select two cases in the catalog: one with \bh{} spins aligned with the orbital angular momentum and another with precessing \bh{s} (\texttt{GT0582} and \texttt{GT0560} cases respectively in the catalog, see next Section). Fig. ~\ref{f:err-analysis} summarizes the accumulated numerical errors in the \gw{} strain, $h(t)$, from combined $\ell=2:6, m=-\ell:\ell$ radiated mode. The left panels show the results for the aligned-spinning case \texttt{GT0582}, and the right panels for the precessing-spin case \texttt{GT0560}.  Top row panel depicts the strain $h(t)$. The middle and bottom panels show accumulated errors in phase and amplitude for each of the available resolutions, four resolutions for \texttt{GT0582} and three for \texttt{GT0560}. For each resolution, the errors are computed against a waveform obtained from Richardson extrapolation to the continuum using the available resolutions.

Table \ref{tbl:err-analysis} summarizes the errors in phase and amplitude are also reported for early inspiral and reference frequency $M\omega = 0.2$ (near merger). The error-analysis is similar to the one reported in \cite{Hinder:2013oqa}.  The mismatches are computed between two finite numerical grid resolutions and finite waveform extraction radius ($R$ in Eq. \ref{eq:Psi4Decom}). These match calculations involve advanced LIGO noise curve and total-mass of BBH scaled at $100\,M_\odot$.

\section{Description of the Catalog}
\label{sec:catalog}

The initial data for each simulation in the catalog are fully characterized by a set of 15 parameters, as described in $\S$~\ref{sec:code}: \bh{} masses $m_1$, $m_2$, spins $\vec \chi_1$, $\vec\chi_2$, momenta $\vec P_1$, $\vec P_2$, and the binary separation $r$. We select code units such that $M = m_1+m_2 = 1$. The waveforms are classified into two main types: 
\emph{Non-precessing} and \emph{Precessing}. Non-precessing waveforms are subdivided into two sub-types: \emph{Non-spinning} if the \bh{s} in the binary are not spinning, and \emph{Aligned-Spin} if their spins are parallel with the orbital angular momentum $\vec L$  (spins of black hole that are anti-aligned and parallel to $\vec L$ are put under the class of aligned-spin).  The precessing waveforms are also subdivided into two sub-types: \emph{Equal Mass} and \emph{Unequal Mass}. Table~\ref{tbl:bbh-survey} summarizes this classification.  

The catalog can be found at  \href{url}{einstein.gatech.edu/catalog}. Each of the 452 waveforms in the catalog have a unique identifier of the form \texttt{GTXXXX}. The catalog is organized by folders. Each folder contains the following information:
\begin{itemize}
\item Initial parameters of \bbh{} system
\item Parameter file of the simulation
\item \bh{} trajectories
\item Mass, spin and gravitational recoil of the final \bh{}
\item Radiated energy, linear momentum and angular momentum
\item $\Psi_4$ decomposed in spin-weighted spherical harmonics with $\ell \le 8$ and different extraction radii 
\item The waveforms are available in \texttt{HDF5} format with meta-data as stated in Ref.~\cite{SchmidtHarry:2016, GalleySchmidt:2016}.
\end{itemize}

Fig~\ref{f:runs} provides a general sense of the parameter space covered by the catalog. The vertical axis in both plots denotes the mass ratio $q$. The plot in the left is for non-spinning and aligned-spin systems. Therefore, the axis in the plane are given in terms of $\vec \chi_{1,2} \cdot \hat L $ in order to capture both the spin magnitude and orientation for each \bh{.} 
The plot on the right in Fig.~\ref{f:runs}  describes the precessing runs. The axis in the plane are given in this case in terms of the $\hat \chi_{1,2} \cdot \hat L $, namely the spin orientation relative to the orbital angular momentum. %\jefe{we need to change the labels in the axis in the figure to $\hat \chi_{1,2} \cdot \hat L $}

\begin{figure}[b!]
		\includegraphics[scale=0.32, trim = 40 20 0 30 ]{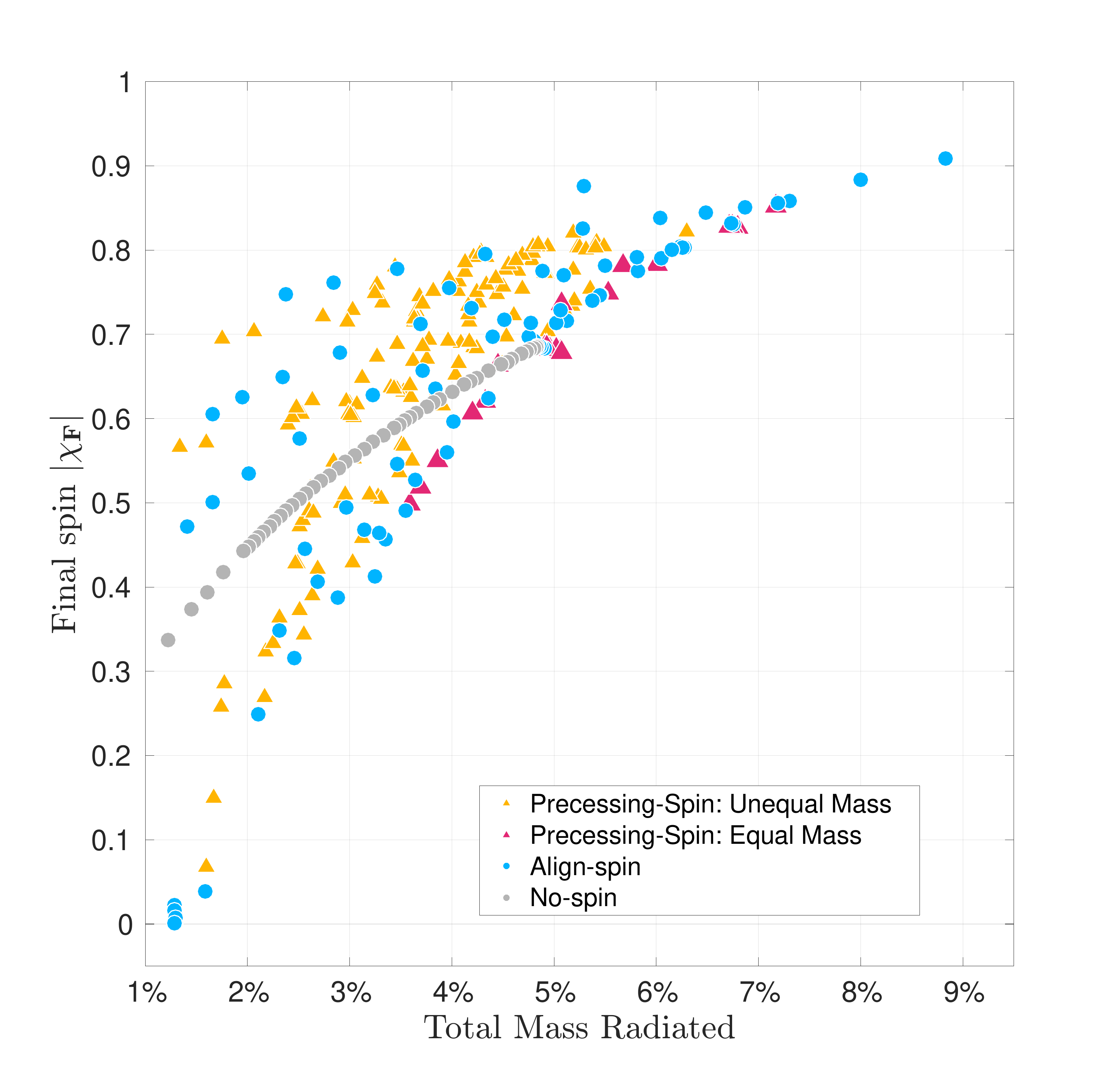}
\caption{Magnitude of the spin of the final \bh{} $|\vec \chi_F|$ as a function of the percentage of total mass radiated, i.e.  $(1- M_F/M)\times 100\%$}
\label{f:final_mass_spin}
\end{figure}

The scatter plot in Fig.~\ref{f:final_mass_spin} shows $|\vec \chi_F|$, the magnitude of the spin of the final \bh{,} as a function of the percentage of total mass radiated, i.e.  $(1- M_F/M)\times 100\%$. Notice that binary systems with high final \bh{} spin radiate the most energy. On the other hand, configurations that leave behind a slowly rotating \bh{} radiated very little.

\begin{table}[h]
  \begin{center} \begin{tabular}{llr}
	  Type    & Sub-type &  Simulations\\
  		\hline \hline 	    
		Non-Precessing       &  Non-Spinning &  49    \\
					       &  Aligned-Spins &   79    \\ 
		Precessing    &  Equal-Mass ($q=1$) &  127    \\
								  &  Unequal-Mass ($q\ne 1$)&  197 
  	\end{tabular}
  \end{center}
  \caption{GT catalog waveform classification}
  \label{tbl:bbh-survey}
\end{table}

\begin{figure}[htb!]
				\includegraphics[scale=0.32, trim = 40 20 0 0 ]{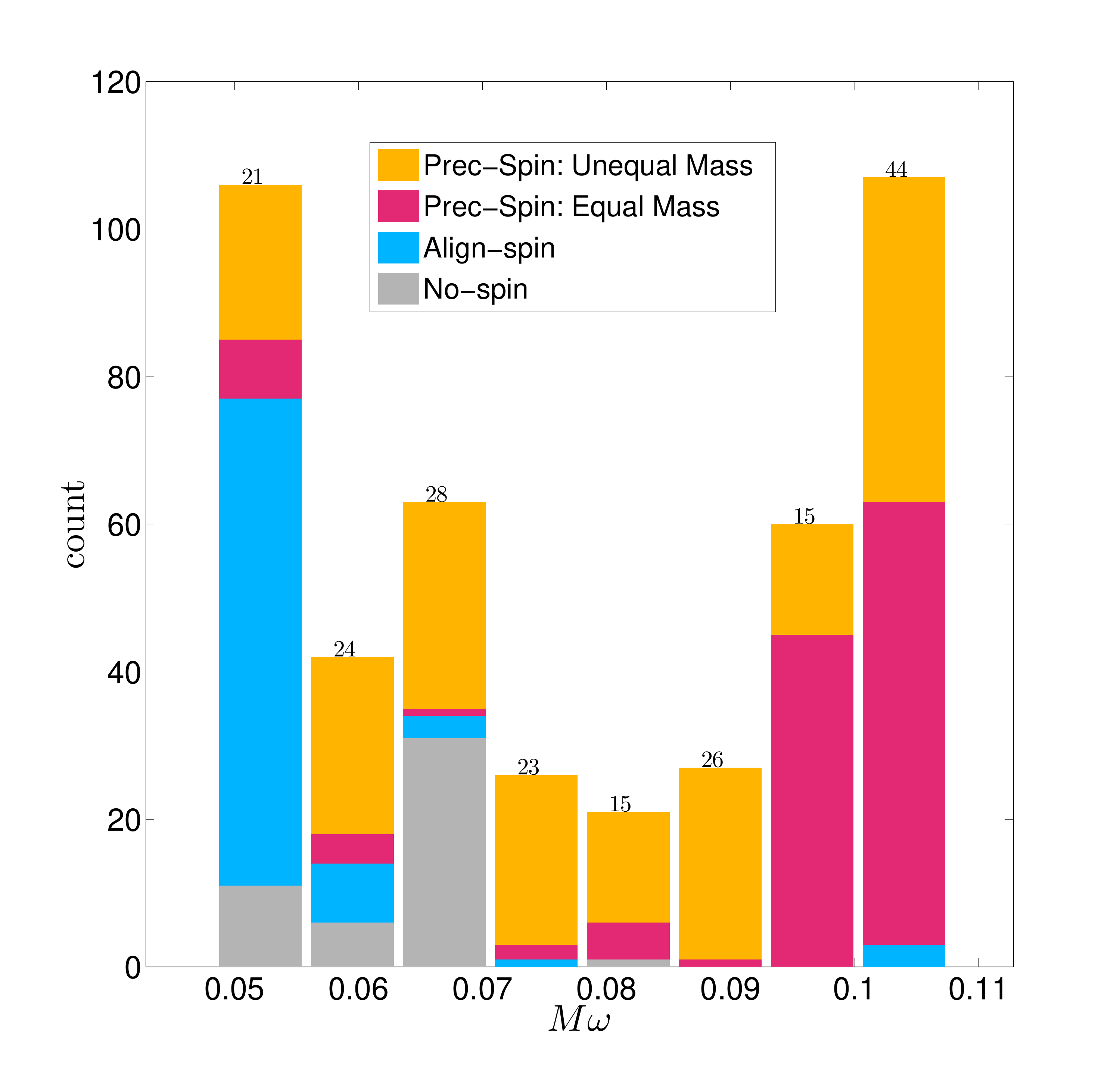}
\caption{The histogram showing the distribution of the $M\omega_{\text{orb}}$. For Advanced LIGO, the x-axis limit scales to $\left[50, 110 \right]\,M_\odot $ as range of minimum total mass $M$. The numbers on the top of each bar corresponds to the total unequal-mass precessing simulations in the stated range. 
}
\label{f:mass_range}
\end{figure}

The histogram in Fig.~\ref{f:mass_range} show the distribution  $M\omega_{\text{orb}}$ where $\omega_{\text{orb}}$ the orbital frequency (half of gravitational-wave frequency). The \nr{} waveforms presented the catalog include the early phase of the simulation that is contaminated with the junk radiation in the initial data. The segment of the waveform with orbital frequencies $\le M\omega_{\text{orb}}$ should hence be ignored. For a given low-frequency cutoff of a \gw{} detector, $f_{\text{min}}$, the waveform can be scaled to a minimum total mass as $M =  k \left( M\omega_{\text{orb} } / f_{\text{min}} \right)$, where $k=3.23\times10^{4}$.  From Fig.~\ref{f:mass_range} it can be inferred the catalog includes a large number of waveforms with less than four \gw{} cycles. These are basically \bbh{} plunges. They are nonetheless useful for studies of quasinormal ringing and gravitational recoil. Waveforms with between five to ten \gw{} cycles are suitable to investigate \bbh{} with massive \bh{} such as GW150914. 

\begin{figure*}[htb!]
	\includegraphics[scale=0.3, trim = 30 20 0 0 ]{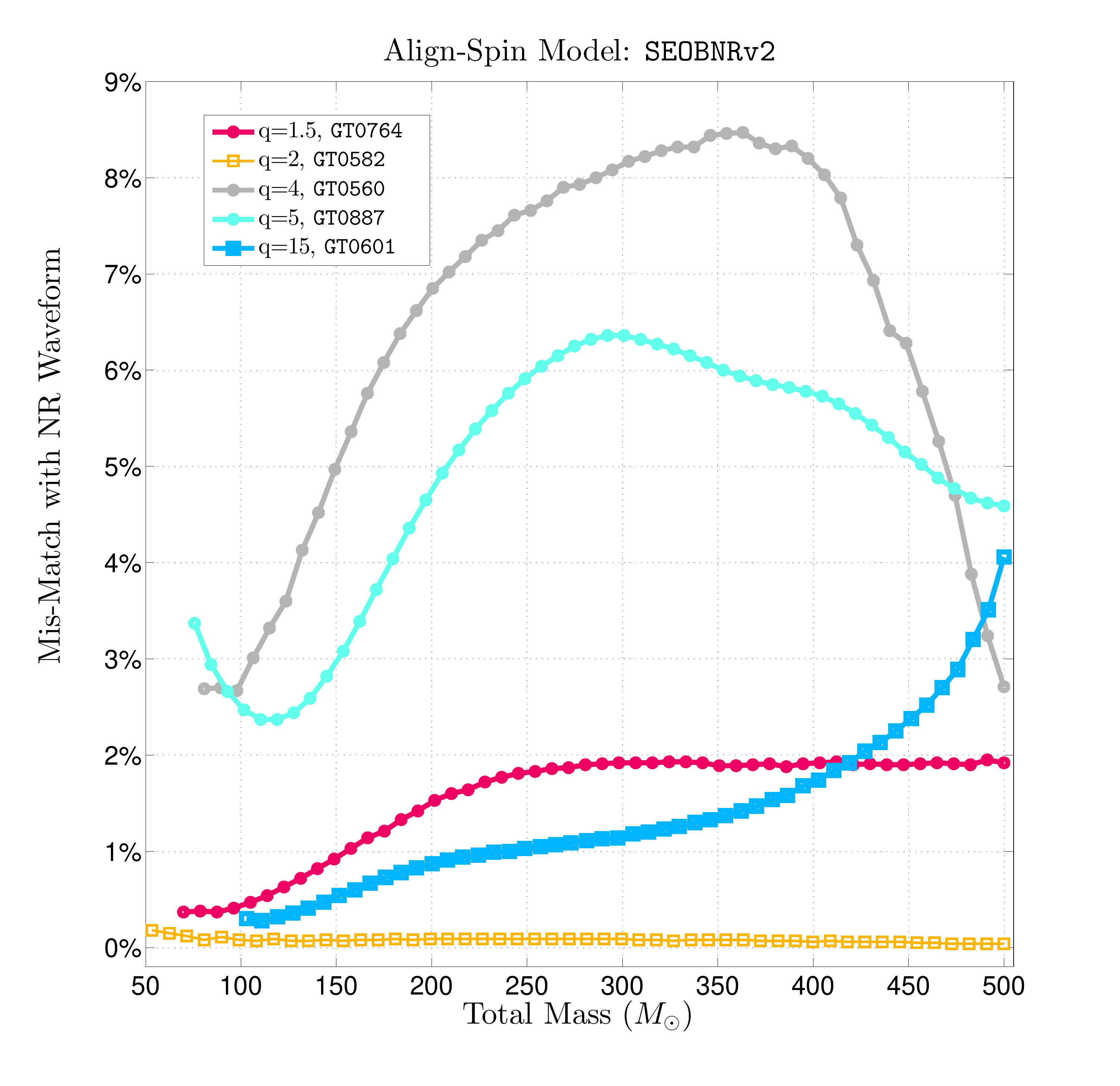}				\includegraphics[scale=0.3, trim = 10 20 40 0 ]{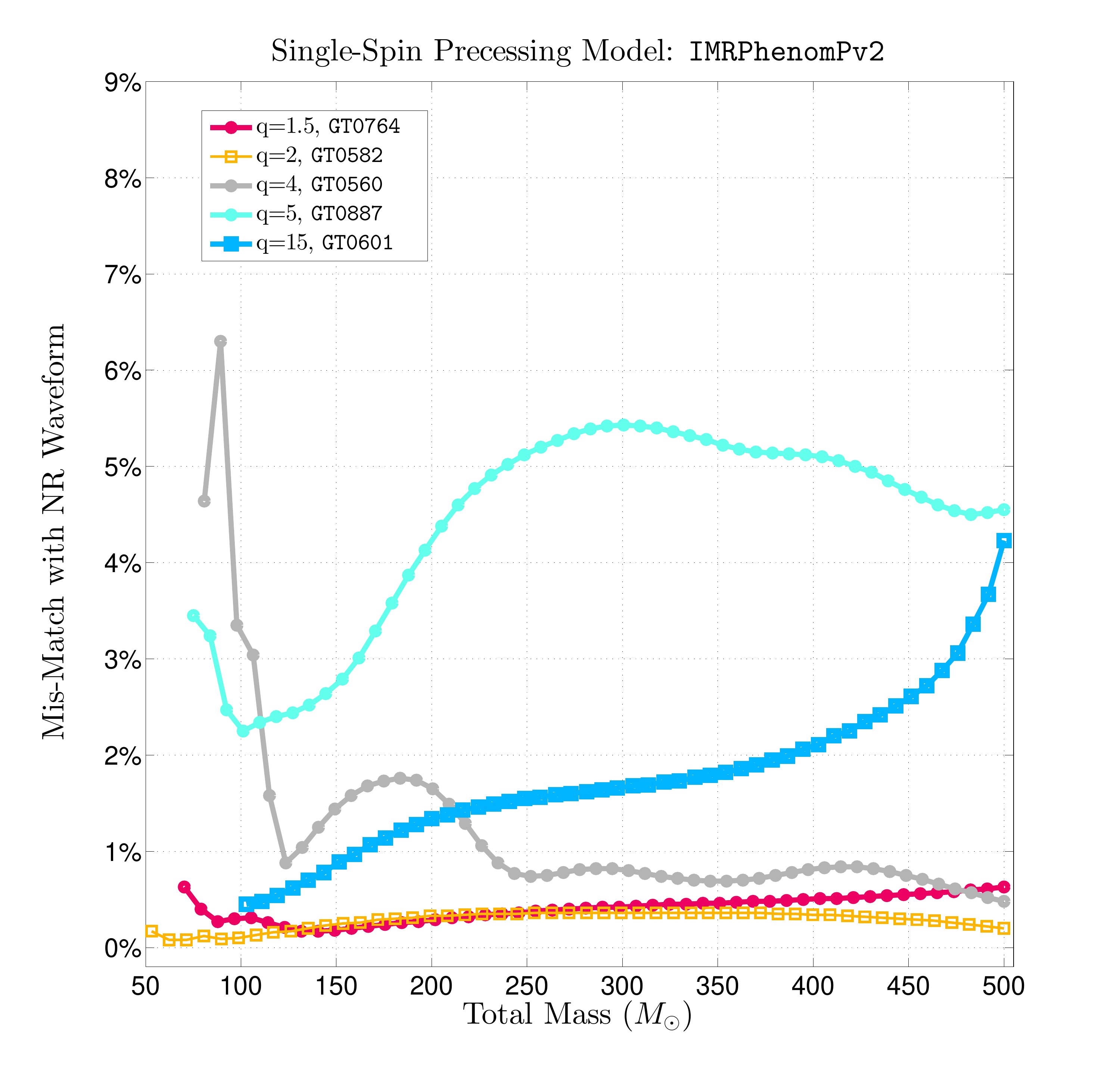}
\caption{Mismatches of \nr{} waveforms in Table~\ref{tbl:matchNR} with approximant GW models.
}
\label{f:match}
\end{figure*}  

Some of the highlights in the catalog are: The largest mass-ratio is $q=15$ for a non-spinning \bbh{} (\texttt{GT0601}), while for precessing \bbh{} $q=8$ (\texttt{GT0886}). The maximum spin for the merging \bh{} is $|\chi_{1,2} |= 0.8$. The most extreme spin for the remnant \bh{} is $|\chi_{F} | = 0.9048$, corresponding to $8.826\%$ of the total-mass $M$ radiated in \gw{s} (\texttt{GT0424})
(see Fig.~\ref{f:final_mass_spin}). The maximum total angular momentum radiated is $\sim 100\%$ for a system of align-spin \bbh{} which results in Schwarzschild-like remnant \bh{} (\texttt{GT0770}). The maximum \gw{} cycles in our simulation corresponds to 27.5 for align-spin  (\texttt{GT0612}) and 21.5 for precessing-spin systems (\texttt{GT0468}).

\section{Comparison With Appproximant Gravitational-Waveform Models}
\label{sec:EOB}

Next we compare a few of the waveforms in the catalog with two recent and well-known approximate waveforms. The binary parameters of the
selected waveforms are given in Table~\ref{tbl:matchNR}, and the corresponding strains $h(t)$ for the two cases ((\texttt{GT0582} and \texttt{GT0560}) are show in Fig.~\ref{f:err-analysis} . The cases were chosen to probe highly distinct regions of parameter space.

The two approximate waveform models we use to compare our \nr{} waveforms are:  
 i) a time-domain model for non-precessing, aligned-spin
systems, derived from the effective-one-body formalism (referred to as {\tt
SEOBNRv2}~\cite{SEOBNRv2, SEOBNRv2ROM:PhysRevD.93.064041}) and ii) a phenomelogical frequency-domain model for
single-spin, precessing systems (referred to as {\tt IMRPhenomPv2}~\cite{IMRPhenomPv2:PhysRevD.93.044007, IMRPhenomPv2, IMRPhenomPv2:PhysRevLett.113.151101}).
Both of these approximate models were used in the detection and parameter-estimation analysis of GW150914.

For each waveform in Table~\ref{tbl:matchNR}, we compute their mismatch with both \texttt{SEOBNRv2} and  \texttt{IMRPhenomPv2}, where the mismatch is given by
\begin{equation}\label{eq:match}
    \mathrm{mismatch} = 1-\max_{t_0,~\phi_0} \frac{(h_1| h_2 )}
    {\sqrt{(h_1|h_1)(h_2|h_2)}},
\end{equation}
where the inner product is given by
\begin{equation}\label{eq:inner_product}
    (h_1|h_2) = 4 \mathrm{Re} \int_{f_\mathrm{min}}^{\infty}
    \frac{\tilde{h_1}(f)\tilde{h_2}^*(f)}{S_h(f)}~{\mathrm d} f\,.
\end{equation}
The maximization in the mismatch (\ref{eq:match}) is over the initial arrival time and phase. In Eq.~(\ref{eq:inner_product}), 
 $S_h(f)$ is the noise spectral density of the detector, and  asterisks
denote complex conjugation.   The integral is evaluated from some minimum
frequency $f_{\mathrm{min}}$, below which there is no appreciable contribution
to the integrand due to the noise spectrum. 
We set as low-frequency cutoff $f_{\mathrm{min}}=30$\,Hz and use a noise spectrum representative of
advanced LIGO in its early configuration.  To evaluate mismatch, both the
waveforms, \nr{} and  the approximant models, are projected to the same optimal sky-location and orientation.

Figure~\ref{f:match} shows the mismatches for the \nr{} waveforms in Table~\ref{tbl:matchNR} with \texttt{SEOBNRv2} and  \texttt{IMRPhenomPv2}. The mismatch is computed for different values of total mass of \bbh{} systems, starting from \bbh{} systems with mass similar to GW150914 to intermediate mass \bbh{} range for current generation of \gw{} detectors. The \nr{} waveform includes all the higher harmonics (as stated in eq. \ref{eq:Psi4Decom}) from $\ell = 2$ to 6; however, the approximant waveform includes only radiated mode $\ell=2$, $m=2$, which will be dominant for the chosen optimal sky-location and orientation.

\begin{table}[h]
  \begin{center}
  \begin{tabular}{p{1.2cm} p{2.1 cm}  p{0.5 cm}  r r}

 	    		ID & Type & $q$ & $\vec \chi_1$ & $\vec\chi_2$     \\

\hline 	     
\hline 	    
 	    		\texttt{GT0764} & prec-spin& 1.5 & (0.6,0,0) & (0,0,0.6) \\ 
				\texttt{GT0582} & aligned-spin & 2 &(0,0,-0.15) & (0,0,0.6) \\ 
				\texttt{GT0560} & prec-spin & 4  & (-0.6,0,0) & (-0.6,0,0) \\ 
				\texttt{GT0887} & prec-spin	&  5 & (0.42, 0, 0.42) & (-0.42, 0, -0.42) \\ 
				\texttt{GT0601} & non-spin& $15$ &$(0,0,0)$ & $(0,0,0)$ 
  	\end{tabular}
  \end{center}
  \caption{GT \bbh{} simulations used for comparison with approximate \gw{} models. The results are shown in fig. \ref{f:match}. } 
  \label{tbl:matchNR}
\end{table}

For the aligned spins with low-mass ratio, both models have a very strong agreement with \nr{} waveform. For the non-spinning \bbh{} with mass-ratio of $q=15$, which represents an astrophysical intermediate-mass ratio inspiral \bbh{} system, both \texttt{SEOBNRv2}  and \texttt{IMRPhenomPv2} have a growing mismatch at high total mass. For such high masses, the signal in LIGO will be dominated by the merger and ringdown of \bbh{}, and radiated modes beyond the dominant becomes important ~\cite{Bustillo:2015qty, London:2014cma}. Both the models only includes the dominant modes (2,2) and thus there is strong mismatch, even at optimal sky-location. 

For the precessing-spin \bbh{} systems, it is expected that \texttt{SEOBNRv2} will show strong inconsistency with \nr{} simulations as the model is tuned only for aligned-spin systems. The max mismatch we report for \texttt{SEOBNRv2} in precessing cases, which happens for a system with mass-ratio $q=4$. In contrast, for the same \nr{} simulation (\texttt{GT0560}),  the precessing spin model \texttt{IMRPhenomPv2} - reports an error up to $6\%$ for lower total mass and drops to less than $1\%$ at higher total mass. Both models agree fairly well with \nr{} simulations for almost equal-mass systems, but for strongly deviate for mass-ratios $q=5$ and above (where higher radiated modes become important).

 \section{Mass and Spin of the Final \bh{} and Phenomenological Fits}
\label{sec:fits}

As mentioned before, included in the \GT{} catalog is information regarding the mass and spin of the final \bh{}. Over the years, several phenomenological formulas have been proposed that connect the properties (mass and spin) of the remnant \bh{} with the initial parameters of the \bh{s} in the binary. In this section, we concentrate on two of such phenomenological formulas: one from  \citet{Healy:2014yta}, referred as RIT, and the other from \citet{Barausse:2012qz}, referred as BR. 

\begin{figure}[htb!]
		\includegraphics[scale=0.57, trim = 10 10 450 0 ]{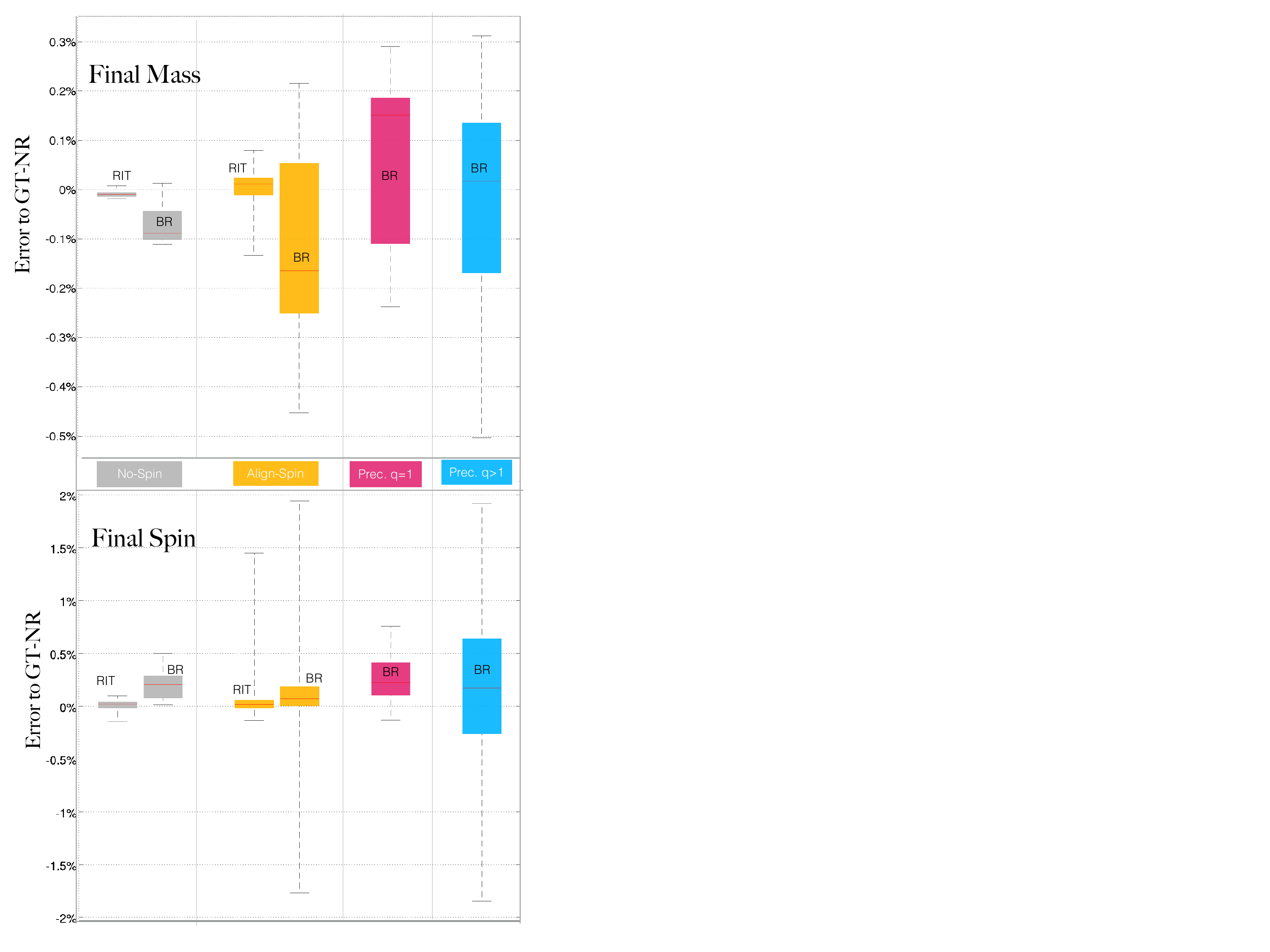}
\caption{Percentage relative errors predicting the mass and spin of the final \bh{} from the \texttt{RIT} and  \texttt{BR} fitting formulas when applied to our catalog. The red line in each box is the median value of the errors. The colored region within each box denote the $25-75$ percentile of relative error in each case.
 }
\label{f:err_final_mass_spin}
\end{figure}

In Figure \ref{f:err_final_mass_spin}, we report the errors the phenomenological formulas incur  in predicting the  mass and spin of the final \bh{.} The percentage errors are organized according to the sub-types in Table~\ref{tbl:bbh-survey}, and they were calculated as $( 1- \text{RIT or BR}/\text{NR})\times 100\%$. Top panels show the errors in the final mass and the bottom for the final spin. The red line in each box is the median value of the errors. On each box, the colored region denotes 75\% of the cases. Notice that, for aligned-spin systems, the spread in errors for the remaining 25\% cases (i.e. cases with the largest errors) is quite significant for both formulas. The \texttt{RIT}, valid for non-spinning and align-spinning \bbh{} systems, has an average discrepancy with our catalog of $0.035\%$ for the remnant mass  and $0.23\%$ for the remnant spin. The \texttt{BR} formula,  valid for all generic \bbh{} configurations, agrees remarkably with all our GT-\bbh{} simulations, and with an average discrepancy of $0.6\%$ for the final mass  and $1.6\%$ for the final spin. A recent paper by the authors \cite{Hofmann:2016yih} improves the  \texttt{BR} formula for stronger agreements with generic \bbh{} \nr{} simulations. 

\section{Conclusion}
\label{sec:conclude}

This paper introduced the \GT{} catalog of \gw{} waveforms consisting of 452 distinct waveforms from more than 600 spin-aligned and precessing \bbh{} simulations with mass ratios of up to $q = 15$. The waveforms expand a moderate number of orbits in the late inspiral, the burst during coalescence, and the ring-down of the final black hole. A significant fraction of the waveforms have enough \gw{} cycles that can be used in improving phenomenological or EOB models.   The waveforms are also useful for tuning the phenomenological formulas describing the remnant black hole.  Most of the waveforms can be used directly in connection with analysis of massive \bbh{} binaries such as GW150914 and for conducting tests of general relativity that require knowledge of both the inspiral and ringdown stages. The \GT{} catalog complements and enhances the catalog recently introduced by the SXS collaborations~\cite{2013PhRvL.111x1104M}.  The GT catalog contains waveforms of the higher modes and will serve as repository of future waveforms, including those from double neutron star and mixed binary mergers. 

%%%%%%%%%%%%%%%%%%%%%%%%%%%%%%%%%%%%%%%%%%%%%%%%%%%%%
\section*{Acknowledgments:} We would like to thank Juan Calderon Bustillo and Sebastian Khan for useful discussions. This work is supported by NSF grants 1505824, 1333360, and XSEDE  TG-PHY120016.

\bibliography{references}

\end{document}